\documentclass[lineno]{biometrika}

\usepackage{amsmath}

\usepackage{times}
\usepackage{bm}
\usepackage{natbib}
\usepackage{graphicx}
\usepackage{caption}
\usepackage{floatrow}

\usepackage[plain,noend]{algorithm2e}

\makeatletter
\renewcommand{\algocf@captiontext}[2]{#1\algocf@typo. \AlCapFnt{}#2} 
\def\@algocf@capt@plain{top}
\renewcommand{\algocf@makecaption}[2]{%
  \addtolength{\hsize}{\algomargin}%
  \sbox\@tempboxa{\algocf@captiontext{#1}{#2}}%
  \ifdim\wd\@tempboxa >\hsize
    \hskip .5\algomargin%
    \parbox[t]{\hsize}{\algocf@captiontext{#1}{#2}}
  \else%
    \global\@minipagefalse%
    \hbox to\hsize{\box\@tempboxa}
  \fi%
  \addtolength{\hsize}{-\algomargin}%
}
\makeatother

\def\T{{ \mathrm{\scriptscriptstyle T} }}
\def\R{{ \mathrm{\scriptscriptstyle R} }}

\def\post{\mbox{\normalfont{\scriptsize post}}}
\def\SUN{\textsc{sun}}

\begin{document}

\jname{}
\jyear{}
\jvol{}
\jnum{}
\accessdate{}


\markboth{Daniele Durante}{Conjugate Bayes for probit regression via unified skew-normals}

\title{Conjugate Bayes for probit regression \\ via unified skew-normals}

\author{Daniele Durante}
\affil{Department of Decision Sciences and Bocconi Institute for Data Science and Analytics, Bocconi University, Via R\"ontgen 1, 20136, Milan, Italy \email{daniele.durante@unibocconi.it} }

\maketitle

\begin{abstract}
Regression models for dichotomous data are ubiquitous in statistics. Besides being useful for inference on binary responses, these methods serve also as  building blocks in more complex formulations, such as density regression,  nonparametric classification and graphical models. Within the Bayesian framework, inference proceeds by updating the priors for the coefficients, typically set to be Gaussians,  with the likelihood induced by probit or logit regressions for the responses. In this updating, the apparent absence of a tractable posterior has motivated a variety of computational methods, including Markov Chain Monte Carlo routines and algorithms which approximate the posterior. Despite being routinely implemented,  Markov Chain Monte Carlo strategies face mixing or time-inefficiency issues in large $p$ and small $n$ studies, whereas approximate routines fail to capture the skewness typically observed in the posterior. This article proves that the posterior  distribution for the probit coefficients has a unified skew-normal kernel, under Gaussian priors. Such a novel result allows efficient Bayesian inference for a wide class of applications, especially in large $p$ and small-to-moderate $n$ studies where state-of-the-art computational methods face notable issues. These  advances are  outlined in a genetic study, and further motivate the development of a wider class of conjugate  priors for probit models along with methods to obtain independent and identically distributed samples from the unified skew-normal posterior.
\end{abstract}


\section{Introduction}
\label{sec.1}
There is a relevant interest in several fields towards learning how the probability mass function of a binary response $y \in \{0;1\}$ varies with a set of observed predictors $x=(x_1, \ldots, x_p)^{\T} \in \Re^p$  \citep[e.g.][]{agrest_2013}. To address this goal, common formulations assume $y$ is a Bernoulli variable whose  probability parameter changes with a linear combination of the predictors under a probit or logit mapping. In the first case $\mbox{pr}(y=1 \mid x,\beta)=\Phi(x^{\T} \beta)$, whereas in the second  $\mbox{pr}(y=1 \mid x,\beta)=\{1+\exp(-x^{\T} \beta) \}^{-1}$, and the goal is to provide inference on $\beta=(\beta_1, \ldots, \beta_p)^{\T} \in \Re^p$. 

Although frequentist inference for the above class of models is well-established \citep[e.g.][]{agrest_2013}, the Bayesian approach has attracted an increasing interest since it allows borrowing information, uncertainty quantification, shrinkage and tractable inference via the posterior distribution for the regression coefficients \citep[e.g.][\S 7.2]{agrest_2013}. Besides this, predictor-dependent models for binary data are also useful building blocks in more complex Bayesian formulations, such as density regression models \citep{rodrig_2011}, additive trees \citep{chipman_2010},  nonparametric classification \citep{ras_2006}, graphical models \citep{Spiegel_1990},    and others. Although these methods provide popular learning procedures,  there are still computational barriers. Indeed, unlike Bayesian regression with Gaussian data, there are no results on the availability of tractable posteriors for $\beta$ in regression models for Bernoulli data, under the common Gaussian priors on the coefficients  \citep[e.g.][]{chopin_2017}.

Motivated by the above issue, several computational methods have been proposed for Bayesian regression with binary response data. Popular routines consider data augmentation strategies relying on hierarchical representations which provide conjugate full conditionals, within a Markov Chain Monte Carlo \citep{Albert_1993, Polson_2013,holmes_2006,fru_2007}. Although being routinely implemented, these methods face poor convergence and mixing in practice, especially for imbalanced datasets \citep{john_2017}. A solution is to consider alternative strategies, including carefully tuned or adaptive Metropolis--Hastings \citep{rob_2001,haario_2001}, as well as more recent generalizations of Hamiltonian Monte Carlo, such as the no u-turn sampler by \citet{hoff_2014}. Both  procedures guarantee computational advantages in large $n$ and moderate $p$ settings compared to data augmentation Markov Chain Monte Carlo. However, when $p$ is large, Metropolis--Hastings has difficulties in exploring the parametric space, whereas Hamiltonian Monte Carlo tends to be expensive  \citep{chopin_2017}. Laplace approximations, variational Bayes and expectation propagation  scale-up computations, but commonly provide Gaussian approximations which affect quality of inference  when the posterior is skewed. This is a common situation in regression for binary data \citep[][]{kuss_2005} and, as will be discussed in the rest of this article, a property which is inherent to the posterior. Refer to \citet{chopin_2017}  for a thorough discussion and comparison among these methods with a specific focus on probit regression. 

Although providing state-of-the-art methods in Bayesian regression for binary response data, the aforementioned strategies are still  sub-optimal compared to situations in which the posterior belongs to a known and tractable class of random variables. Indeed, this result could facilitate the calculation of several quantities relevant to posterior inference, without relying on Monte Carlo methods. This article proves that when the focus is on probit regression models under Gaussian priors for the coefficients, the posterior belongs to the  class of unified skew-normal distributions \citep{arellano_2006}. These variables already appeared in probit models to obtain  flexible link functions via skewed  latent data, instead of Gaussian ones \citep[e.g.][]{bazan_2006}. However, this is a different contribution compared to the one in the present article.

To the best of the author's knowledge, the above result is not available in the literature, but can provide important advances. In fact, as discussed in this article, the unified skew-normal posterior guarantees closure properties  \citep{arellano_2006} in addition to explicit formulas for the marginal, joint and conditional posteriors along with predictive distributions and marginal likelihoods for model selection. These quantities involve cumulative distribution  functions $\Phi_n(\cdot)$ of $n$-variate Gaussians, and hence can be efficiently evaluated  for small-to-moderate $n$ studies via recent minimax tilting methods \citep{botev_2017}. The contribution by \citet{botev_2017} is additionally useful to obtain independent samples from the posterior exploiting a representation of the unified skew-normal via a linear combination of $p$-variate Gaussians and $n$-variate truncated Gaussians. Such results are valid for any dimension and provide, {\em per se}, key methodological advances which motivate future theoretical studies and facilitate formal understanding of the skewness typically observed in the posterior. However, the associated inference strategies require evaluation of $\Phi_n(\cdot)$ or sampling from $n$-variate truncated Gaussians, and hence are of practical usefulness in small-to-moderate $n$ studies with, typically, few hundreds of units and any, even huge, $p$. This scenario is the most challenging for current Markov Chain Monte Carlo routines  \citep[e.g][]{chopin_2017} and, in fact, the methods for posterior inference arising from the new unified skew-normal result significantly improve state-of-the-art algorithms in these situations, thus covering also an important computational gap; refer to \S \ref{sec.24} for a more detailed discussion on these aspects. 

\section{Posterior inference in probit regression via unified skew-normals}
\label{sec.2}
\subsection{The unified skew-normal distribution}
\label{sec.21}
Before deriving the unified skew-normal posterior induced by Gaussian priors for the $\beta$ coefficients in a probit regression, let us first introduce the unified skew-normal distribution. Recalling  \citet{arellano_2006}, this random variable unifies different generalizations of the multivariate skew-normal  $z \sim \textsc{sn}_p(\xi,\Omega,\alpha)$ \citep{azza_1996} whose density $2 \phi_p(z-\xi; \Omega) \Phi\{\alpha^\T \omega^{-1}(z-\xi) \}$ is obtained by modifying the one of a $p$-variate Gaussian $N_{p}(\xi, \Omega)$ with the  cumulative distribution function of a standard normal evaluated at $\alpha^\T \omega^{-1}(z-\xi)$, where $ \omega$ is a $p \times p$ diagonal matrix containing the square root of the diagonal elements in $\Omega$. This strategy  introduces skewness in $N_{p}(\xi, \Omega)$ controlled by  $\alpha=(\alpha_1, \ldots, \alpha_p )^{\T} \in \Re^p$, with $\xi=(\xi_1, \ldots, \xi_p)^{\T}  \in \Re^p$ and $\Omega$ driving location and variability, respectively \citep[e.g.][]{arellano_2006}. Indeed, when $\alpha=0_p$ the multivariate skew-normal coincides with $N_{p}(\xi, \Omega)$, whereas, setting $p=1$, leads to a univariate skew-normal $\textsc{sn}(\xi,\omega^2,\alpha)$  \citep{azza_1985}. 

Motivated by the success of the above formulation in different studies \citep[e.g.][]{azza_1999}, several extensions have been proposed to incorporate additional properties. Two important generalizations are obtained by adding an additional parameter $\gamma$ in $ \Phi\{\alpha^\T \omega^{-1}(z-\xi) \}$ to develop the multivariate extended skew-normal \citep{arnold_2000,arnold_2002}, and by allowing the skewness-inducing mechanism to be multivariate, thus providing the closed skew-normal family \citep{gupta_2004,gonz_2004} which includes a skewness matrix $\Delta \in \Re^{p \times n}$ and an $n \times n$ full-rank scale $\Gamma$ in $\Phi_n({\cdot})$. Besides increasing flexibility, these extensions allow closure properties for marginals, conditionals and joint distributions, thus providing a general class. \citet{arellano_2006} unify the above generalizations within a single and tractable unified skew-normal representation, obtaining the following density function
\begin{eqnarray}
\phi_p(z -\xi; \Omega) \frac{\Phi_n\{\gamma+\Delta^\T \bar{\Omega}^{-1} \omega^{-1}(z-\xi); \Gamma-\Delta^{\T}\bar{\Omega}^{-1}\Delta \}}{\Phi_n(\gamma;\Gamma)},
\label{eq1}
\end{eqnarray}
for $z \sim \SUN_{p,n}(\xi,\Omega,\Delta,\gamma,\Gamma)$. In \eqref{eq1}, $\phi_p(z -\xi; \Omega)$ denotes the density of a $p$-variate Gaussian with expectation $\xi=(\xi_1, \ldots, \xi_p)^{\T}  \in \Re^p$,  and $p\times p$ variance-covariance matrix $\Omega=\omega\bar{\Omega}\omega$ obtained via the quadratic combination between a correlation matrix $\bar{\Omega}$ and a diagonal matrix $ \omega$ containing the square root of the diagonal elements in $\Omega$. The quantities $\Phi_n\{\gamma+\Delta^\T \bar{\Omega}^{-1} \omega^{-1}(z-\xi); \Gamma-\Delta^{\T}\bar{\Omega}^{-1}\Delta \}$ and $\Phi_n(\gamma;\Gamma)$ denote instead the cumulative distribution functions of the multivariate Gaussians $N_n(0_n, \Gamma-\Delta^{\T}\bar{\Omega}^{-1}\Delta)$ and $N_n(0_n, \Gamma)$, evaluated at $\gamma+\Delta^\T \bar{\Omega}^{-1} \omega^{-1}(z-\xi)$ and $\gamma$, respectively,  with the $p \times n$ matrix $\Delta$ having the main effect on skewness. In fact, when  $\Delta$ is zero, \eqref{eq1} coincides with the density of a $N_{p}(\xi, \Omega)$. The vector $\gamma \in \Re^n$ adds additional flexibility in departures from normality, consistent with the  multivariate extended skew-normal. Refer to \citet{arellano_2006} and \citet[][\S 7.1.2]{azzalini_2013} for details. 

It shall be noticed that \citet{arellano_2006} add a further condition which restricts the $(n+p) \times (n+p)$ matrix $\Omega^*$, having blocks $\Omega_{[11]}^*=\Gamma$, $\Omega_{[22]}^*=\bar{\Omega}$ and $\Omega_{[21]}^*=\Omega_{[12]}^{*\T}=\Delta$, to be a full-rank correlation matrix. As will be clarified in \S \ref{sec.22}, this identifiability restriction is not required in the Bayesian setting. In fact, the parameters of the unified skew-normal posterior for the coefficients $\beta$ are functions of the observed data and the pre-specified hyperparameters of the Gaussian prior, thus avoiding identifiability issues. Nonetheless, such a parameterization will be maintained in the  article to inherit the classical results of the unified skew-normal distribution and to ensure identifiability of the prior, when the findings for the Gaussian case will be generalized to the entire class of  unified skew-normals priors. Sections \ref{sec.22}--\ref{sec.23} prove that the posterior for the $\beta$ coefficients in a probit model with Gaussian priors is a unified skew-normal and study the consequences of this novel finding in posterior inference. 

\subsection{Unified skew-normal posterior for Bayesian probit regression with Gaussian priors}
\label{sec.22}
To introduce the reader to the general case consisting of $n$ observations from a probit model with Gaussian prior $\pi(\beta)=  \phi_{p}(\beta-\xi; \Omega)$, let us first consider a simple setting with a single data point $y$ and one covariate $x$, such that $(y \mid x, \beta) \sim \mbox{Bern}\{\Phi(x \beta)\}$ and $\beta \sim N(0,1)$. Although this scenario is uncommon in practice, it provides key intuitions on the role of $x \in \Re$ and $y \in \{0;1\}$ in driving departures from normality in the posterior distribution.  Indeed, consistent with Lemma \ref{lemma1}, $(\beta \mid y, x)$ has a unified skew-normal posterior when $\pi(\beta)=\phi(\beta)$. See Appendix A for proofs.

\begin{lemma}
Let $(y \mid x, \beta) \sim \mbox{\normalfont Bern}\{\Phi(x \beta)\}$ and set $\pi(\beta)=\phi(\beta;1)=\phi(\beta)$, then $(\beta \mid y, x) \sim \SUN_{1,1}\{0,1, (2y-1)x(x^{2}+1)^{-1/2},0,1\}$, for every $x \in \Re$ and $y \in \{0;1\}$. 
\label{lemma1}
\end{lemma}

\begin{figure}[t]
\captionsetup{font={small}}
\includegraphics[height=6.2cm]{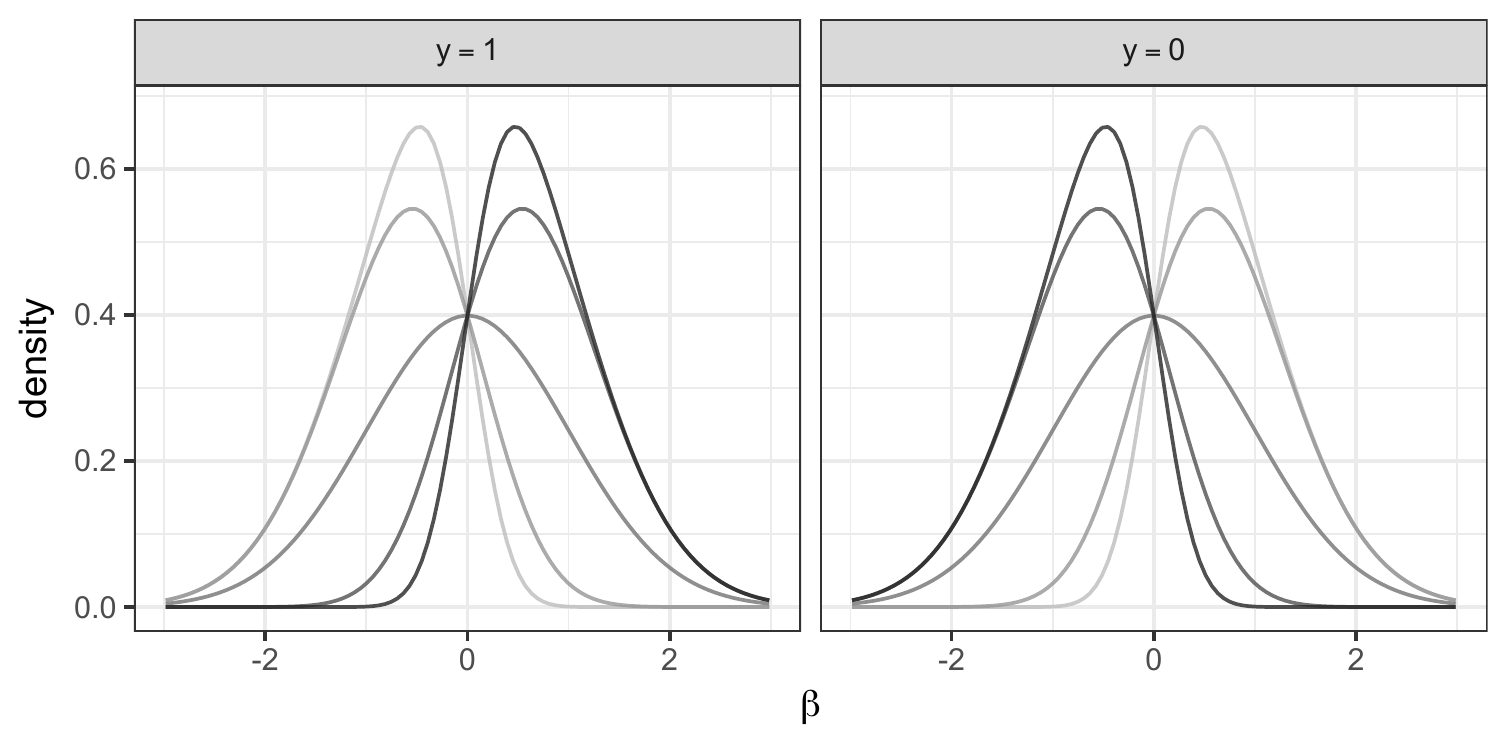}
\caption{Density of a $\SUN_{1,1}\{0,1,(2y-1)x(x^{2}+1)^{-1/2},0,1\}$ posterior for $\beta$, under varying $x$ and $y$. Colors range from light grey to dark grey as $x \in (-3,-1{\cdot}5,0,1{\cdot}5,3)$ goes from $-3$ to $3$.}
\label{f1}
\end{figure}

Figure \ref{f1} provides the density function of the unified skew-normal posterior for $\beta$ in the illustrative example, under different combinations of $x$ and $y$. As expected, $(2y-1)x(x^{2}+1)^{-1/2}$ controls  skewness. Indeed, the higher $|x|$ the more skewness is observed in the posterior. This skewness is either positive or negative depending on the sign of $(2y-1)x$.   To better understand this result, note that the unified skew-normal in Lemma \ref{lemma1}  coincides with a basic $\textsc{sn}\{0,1,(2y-1)x\}$.

The above results apply more generally to independent response data $y_1, \ldots, y_n$ from a probit model $(y_i \mid x_i, \beta) \sim \mbox{Bern}\{\Phi(x_i^{\T} \beta) \}$, for $i=1, \ldots, n$, where $x_i=(x_{i1}, \ldots, x_{ip})^{\T} \in \Re^p$ denotes the vector of covariates for unit $i$ and $\beta=(\beta_1, \ldots, \beta_p)^{\T}\in \Re^p$ the associated coefficients.  Indeed, based on Theorem \ref{teo1}, when $\beta$ has a Gaussian prior $\beta \sim N_p(\xi,\Omega)$ with mean $\xi \in \Re^p$ and full-rank variance-covariance matrix $\Omega=\omega\bar{\Omega}\omega$, the posterior coincides with a unified skew-normal. 

\begin{theorem}
If $y=(y_1, \ldots, y_n)^{\T}$ comprises conditionally independent binary response data from a probit model $(y_i \mid x_i, \beta) \sim \mbox{\normalfont Bern}\{\Phi(x_i^{\T} \beta)\}$, for $i=1, \ldots, n$, and $\beta \sim N_p(\xi,\Omega)$, then 
\begin{eqnarray}
(\beta \mid y, X) \sim \SUN_{p,n}(\xi_{\post},\Omega_{\post},\Delta_{\post} ,\gamma_{\post}, \Gamma_{\post}),
\label{eq2}
\end{eqnarray}
with the posterior parameters defined as a function of the data and the prior parameters via $$\xi_{\post}= \xi, \quad \Omega_{\post}= \Omega, \quad \Delta_{\post}= \bar{\Omega}\omega D^{\T}s^{-1}, \quad  \gamma_{\post}= s^{-1}D\xi, \quad \Gamma_{\post}= s^{-1}(D \Omega D^{\T}+\mbox{ I}_n)s^{-1},$$ for every $n\times p$ data matrix $D=\mbox{\normalfont diag}(2y_1-1, \ldots, 2y_n-1)X$ and any $n \times n$ positive diagonal matrix $s=\mbox{\normalfont  diag}\{(d_1^{\T}\Omega d_1+1)^{1/2}, \ldots, ({d_n^{\T}\Omega d_n+1})^{1/2}\}$. The generic vector $d_i^{\T}$ denotes instead the $i$-th row of $D$, whereas $X$ is the design matrix and $I_n$ the $n \times n$ identity matrix.
\label{teo1}
\end{theorem}

Adapting equation \eqref{eq1} to the results in Theorem \ref{teo1}, it can be immediately noticed, after minor mathematical simplifications, that the density function of the unified skew-normal posterior is 
\begin{eqnarray}
\pi(\beta \mid y,X)=\phi_p(\beta - \xi; \Omega) \frac{\Phi_n\{s^{-1}D\beta;(s s^{\T})^{-1}\}}{\Phi_n\{s^{-1}D\xi; s^{-1}(D \Omega D^{\T}+ I_n )s^{-1}\}},
\label{eq3}
\end{eqnarray}
where $\Phi_n\{s^{-1}D\xi; s^{-1}(D \Omega D^{\T} + I_n)s^{-1} \}$ defines the normalizing constant. To clarify  the role of the prior parameters $\xi$ and $\Omega$, along with that of the data $y$ and $X$, let us consider a constructive representation of the posterior. In particular, adapting known results  from   unified skew-normals  \citep[][\S 7.1.2]{azzalini_2013} to the specific posterior in Theorem \ref{teo1}, it can be shown that $(\beta \mid y, X)$ has the stochastic representation in Corollary \ref{cor1}.

\begin{corollary}
If  $(\beta \mid y, X)$ has the unified skew-normal distribution from Theorem \ref{teo1}, then
\begin{eqnarray}
(\beta \mid y, X)\stackrel{\mbox{\scriptsize \normalfont d}}{=} \xi+\omega\{V_0+ \bar{\Omega}\omega D^{\T}(D \Omega D^{\T}+I_n)^{-1}sV_1\}, \qquad (V_0 \perp V_1),
\label{eq4}
\end{eqnarray}
with $V_0 \sim N_{p}\{0_p,\bar{\Omega}-\bar{\Omega}\omega D^{\T}(D \Omega D^{\T}+I_n)^{-1}D \omega\bar{\Omega}\}$ and $V_1$ from a zero mean $n$-variate truncated normal with covariance matrix $s^{-1}(D \Omega D^{\T}+ I_n)s^{-1}$ and truncation below $- s^{-1}D\xi$.  
\label{cor1}
\end{corollary}

Based on \eqref{eq4}, $\xi$ has a main role on the location, but has also an effect in controlling departures from normality  since it appears in the truncation $s^{-1}D\xi$. Instead, the prior variance-covariance matrix $\Omega$ mainly affects scale, via $\omega$, and posterior dependence among the $\beta$ parameters, while contributing also to the weight matrix assigned to the multivariate truncated Gaussian $V_1$, along with its variability. Finally, the data in $D$ play a major role in controlling departures from normality. Indeed, if $D$ has elements close to $0$, the multivariate truncated Gaussian $V_1$ has a negligible importance compared to the multivariate Gaussian $V_0$ in  \eqref{eq4}.

\subsection{Inference, prediction and variable selection under unified skew-normal posteriors}
\label{sec.23}
A general primary focus in Bayesian regression studies is on marginal posteriors $(\beta_j \mid y, X)$, for $j=1, \ldots, p$, their associated moments and more complex functionals including measures of posterior dependence along with  credible intervals or regions. A fundamental property of unified skew-normals, which potentially facilitates this type of inference, is that such a class of  variables is closed under marginalization, linear combinations and conditioning \citep{arellano_2006, azzalini_2013}. In particular, adapting the derivations in  \citet{arellano_2006} to  Theorem \ref{teo1}, each marginal posterior  is still from a unified skew-normal for every  $\beta_j$, $j=1, \ldots, p$. More specifically, $(\beta_j \mid y, X)\sim  \SUN_{1,n}(\xi_{\post j},\Omega_{\post jj},\Delta_{\post j} ,\gamma_{\post}, \Gamma_{\post})$ with $\Delta_{\post j}$ denoting the $j$-th row of $ \bar{\Omega}\omega D^{\T}s^{-1}$,  $\xi_{\post j}$ the $j$-th element of the prior mean vector $\xi$, $\Omega_{\post jj}$ the entry $[jj]$ in $\Omega$, whereas $\gamma_{\post}$ and $\Gamma_{\post}$ coincide with those already defined in Theorem \ref{teo1}. A similar result holds also for sub-vectors of coefficients $(\beta_{\mathcal{J}} \mid y, X)$, with $\mathcal{J}\subset\{1; \ldots; p\}$,  linear combinations $(a+A^{\T}\beta \mid  y, X)$, and conditional posteriors $(\beta_{\mathcal{J}} \mid y, X,\beta_{\mathcal{J}^*})$, with $\mathcal{J} \subset \{1; \ldots; p\}$, $\mathcal{J}^* \subset \{1; \ldots; p\}$, and $\mathcal{J} \cap \mathcal{J}^*=\emptyset$. Refer to \citet[][]{azzalini_2013} for details to obtain the parameters of these unified skew-normals  from simple transformations of those in  Theorem \ref{teo1}. Note that some linear combinations of $\beta$, such as $x^{\T} \beta$, are of particular interest. 

The aforementioned results facilitate graphical representation of marginal or joint posteriors, along with calculation of posterior moments and credible intervals for the probit coefficients via one-dimensional integrals involving marginal posterior densities. This can be done via numerical integration \citep[e.g.][\S 9]{quart_2010} whenever it is possible to evaluate $\Phi_n(\cdot)$ with efficiency and accuracy. When the focus is on posterior moments, another solution is to obtain such quantities via direct derivation of the moment generating function. Indeed, adapting the result in \citet{arellano_2006}  to   \eqref{eq2} a similar strategy can be considered when studying the functionals of the unified skew-normal posterior, provided that $(\beta \mid y, X)$ has moment generating function
\begin{eqnarray}
M(t)=\exp(\xi^{\T}t+0{\cdot}5 t^{\T} \Omega t)\frac{\Phi_n\{ s^{-1}D\xi +s^{-1}D\Omega t;s^{-1}(D \Omega D^{\T}+I_n)s^{-1}\}}{\Phi_n\{s^{-1}D\xi  ; s^{-1}(D \Omega D^{\T}+I_n)s^{-1} \}}, \quad (t \in \Re^p). \qquad
\label{eq5}
\end{eqnarray}
Exploiting \eqref{eq5} and adapting the derivations in \citet{azzalini_2010} to the unified skew-normal in Theorem \ref{teo1}, the posterior expectation of $\beta$ can be explicitly calculated as
\begin{eqnarray}
E(\beta \mid y, X)=\xi+\frac{1}{\Phi_n\{s^{-1}D\xi; s^{-1}(D \Omega D^{\T}+ I_n )s^{-1}\}}\Omega D^{\T}s^{-1}\eta,
\label{eq6}
\end{eqnarray}
where $\eta$ represents an $n \times 1$ vector whose generic $i$-th component is equal to $\phi(\bar{\gamma}_i )\Phi_{n-1}(\bar{\gamma}_{-i}-\bar{\Gamma}_{-i}\bar{\gamma}_i, \bar{\Gamma}_{-i,-i}-\bar{\Gamma}_{-i}\bar{\Gamma}_{-i}^{\T})$, with $\bar{\gamma_i} $ and $\bar{\gamma}_{-i}$ denoting the $i$-th element of $s^{-1}D\xi=\gamma_{\post}$ and the $(n-1) \times 1$ vector obtained by removing the $i$-th entry in $\gamma_{\post}$, respectively. Similarly, $\bar{\Gamma}_{-i,-i}$ defines the sub-matrix of $s^{-1}(D \Omega D^{\T}+ I_n )s^{-1}=\Gamma_{\post}$ without the $i$-th row and column, whereas $\bar{\Gamma}_{-i}$ is the $i$-th column of $\Gamma_{\post}$ with the $i$-th row element removed. Computing the expectation via \eqref{eq6} is more efficient than numerical integration since it requires calculation of $n+1$ cumulative distribution functions, which is typically much less than the number of evaluations of $\Phi_{n}(\cdot)$ required in numerical integration of marginal posteriors. However, as noticed in  \citet{gupta_2013}, obtaining expressions for higher-order marginal and joint moments via direct derivation of \eqref{eq5} requires tedious  calculations, thus motivating Monte Carlo methods based on samples from the posterior, as discussed in  \S \ref{sec.24}. Refer also to  \citet{gupta_2013} and \citet{azzalini_2010} for an expression of the  variance-covariance matrix and the cumulative distribution function of a generic unified skew-normal. Both quantities, appropriately computed under the parameters in Theorem \ref{teo1}, are useful in posterior inference, especially for credible intervals or regions.

Although inference on the posterior distribution of $\beta$ is often of interest,  prediction of a future response $y_{\mbox{\scriptsize new}} \in \{0;1\}$ given the associated covariates $x_{\mbox{\scriptsize new}}\in \Re^p$ and the current data $(y, X)$ is a primary goal in applications of probit models to classification. Within the Bayesian framework, this task requires the derivation of the posterior predictive distribution $(y_{\mbox{\scriptsize new}} \mid y, X, x_{\mbox{\scriptsize new}})$, which is simply a Bernoulli having parameter $\mbox{pr}(y_{\mbox{\scriptsize new}}=1 \mid y, X, x_{\mbox{\scriptsize new}})=\int\Phi(x_{\mbox{\scriptsize new}}^{\T}\beta)\pi(\beta \mid y, X) \mbox{d} \beta$, in the binary case. According to Corollary \ref{cor2}, this probability parameter is available in explicit form.

\begin{corollary}
If $(y_i \mid x_i, \beta) \sim \mbox{\normalfont Bern}\{\Phi(x_i^{\T} \beta)\}$, for $i=1, \ldots, n$, and $\beta \sim N_p(\xi,\Omega)$, then
\begin{eqnarray}
\mbox{\normalfont  pr}(y_{\mbox{\scriptsize \normalfont new}}=1 \mid y, X, x_{\mbox{\scriptsize \normalfont  new}})&=&\frac{\Phi_{n+1}\{s_{\mbox{\normalfont  \scriptsize new}}^{-1} D_{\mbox{\scriptsize \normalfont new}}\xi ; s_{\mbox{\scriptsize \normalfont new}}^{-1}(D_{\mbox{\scriptsize \normalfont new}} \Omega D_{\mbox{\scriptsize \normalfont new}}^{\T}+I_{n+1}) s_{\mbox{\scriptsize \normalfont new}}^{-1}\}}{\Phi_n\{s^{-1} D\xi ; s^{-1}(D \Omega D^{\T}+I_{n}) s^{-1} \}}, 
\label{eq7}
\end{eqnarray}
with $D_{\mbox{\scriptsize  \normalfont  new}}$ representing the $(n+1) \times p$ matrix obtained by adding a last row $d_{\mbox{\scriptsize  \normalfont  new}}^{\T}=x_{\mbox{\scriptsize  \normalfont  new}}^{\T}$ to $D$, whereas $s_{\mbox{\scriptsize \normalfont new}}=\mbox{\normalfont  diag}\{(d_1^{\T}\Omega d_1+1)^{1/2}, \ldots, (d_n^{\T}\Omega d_n+1)^{1/2}, (d_{\mbox{\scriptsize  \normalfont  new}}^{\T}\Omega d_{\mbox{\scriptsize  \normalfont  new}}+1)^{1/2}\}$. 
\label{cor2}
\end{corollary}

An advantage of \eqref{eq7}, compared  to Markov Chain Monte Carlo strategies  \citep[e.g.][]{Albert_1993, holmes_2006, fru_2007, Polson_2013}, is that prediction does not require Monte Carlo integration for $\int\Phi(x_{\mbox{\scriptsize new}}^{\T}\beta)\pi(\beta \mid y, X) \mbox{d} \beta$ via sampling of $\beta$ from the posterior, and hence the computational burden does not depend on $p$.  As will be discussed in \S \ref{sec.24}, this result is especially useful in large $p$ and small-to-moderate $n$ studies. 

The above derivations  are further helpful in obtaining explicit methods to perform Bayesian selection among models $\mathcal{M}_1, \ldots, \mathcal{M}_K$ characterizing, in general, different subsets $\mathcal{J}_1, \ldots, \mathcal{J}_K$ of covariates  entering the linear predictor. Although there are different strategies for model selection \citep[e.g.][]{oha_2009}, the general  approach formally defines prior probabilities $\mbox{pr}(\mathcal{M}_1), \ldots, \mbox{pr}(\mathcal{M}_K)$ for the set of models, and subsequently ranks them via the posterior probabilities $\mbox{pr}(\mathcal{M}_k \mid y, X) \propto \mbox{pr}(\mathcal{M}_k)\int \mbox{pr}(y \mid  \mathcal{M}_k,X, \beta_{\mathcal{J}_k})\pi( \beta_{\mathcal{J}_k} \mid \mathcal{M}_k) \mbox{d}\beta_{\mathcal{J}_k}$, $k=1, \ldots, K$ \citep[e.g.][]{forte_2018,chip_2001}. Clearly, the major issue in this task is the calculation of $ \int \mbox{pr}(y \mid  \mathcal{M}_k, X, \beta_{\mathcal{J}_k})\pi(\beta_{\mathcal{J}_k} \mid \mathcal{M}_k)\mbox{d}\beta_{\mathcal{J}_k}$ which may be intractable in the absence of conjugacy, thus requiring Monte Carlo integration or  approximations \citep[e.g.][]{kass_1995}. This procedure can be implemented  also in probit models leveraging the methods  in \S  \ref{sec.1}, but inference and computational performance face the issues previously discussed. Corollary \ref{cor3} provides instead an explicit formula for the marginal likelihood in probit models with Gaussian priors, which can be easily evaluated, especially in large $p$ and small-to-moderate $n$ settings of interest in such studies.

\begin{corollary}
Let $ \mathcal{M}_k$ denote the probit regression model for $y_1, \ldots, y_n$ including only the covariates with indices in the subset $\mathcal{J}_k \subset \{1;\ldots;p\}$  and assume $(\beta_{\mathcal{J}_k}\mid\mathcal{M}_k) \sim N_{p_k}(\xi_k,\Omega_k)$, with $p_k=|\mathcal{J}_k|$ and  $\beta_{\mathcal{J}_k} \in \Re^{p_k}$ the probit coefficients for the covariates in model $ \mathcal{M}_k$, then
\begin{eqnarray}
\int \mbox{\normalfont pr}(y \mid  \mathcal{M}_k,X, \beta_{\mathcal{J}_k})\pi( \beta_{\mathcal{J}_k} \mid \mathcal{M}_k) \mbox{d}\beta_{\mathcal{J}_k}=\Phi_n\{s_k^{-1}D_k\xi_k;s_k^{-1}(D_k \Omega_k D_k^{\T}+I_n) s_k^{-1} \}, 
\label{eq8}
\end{eqnarray}
for every model $\mathcal{M}_k$, $k=1, \ldots, K$, where $D_k= \mbox{\normalfont diag}(2y_1-1, \ldots, 2y_n-1)X_k \in \Re^{n \times p_k}$, $s_k=\mbox{\normalfont  diag}\{(d_{1k}^{\T}\Omega_k d_{1k}+1)^{1/2}, \ldots, (d^{\T}_{nk}\Omega_k d_{nk}+1)^{1/2}\} \in \Re^{n \times n}_{+}$, and $X_k \in  \Re^{n \times p_k}$ denoting the $n \times p_k$  design matrix of covariates with indices in $\mathcal{J}_k${.}
\label{cor3}
\end{corollary}

Equation \eqref{eq8} is additionally useful to compute Bayes factors \citep[e.g.][]{kass_1995} and to perform Bayesian model averaging \citep[][]{hoe_1999} without sampling from the posterior. 

\subsection{Computational considerations and sampling procedures}
\label{sec.24}

All the inference methods outlined in \S \ref{sec.23} can, in principle, proceed via direct strategies without sampling from the posterior, thus improving the available procedures in large $p$ applications. The only  barrier, which is relevant for  a large $n$, is evaluation of $\Phi_{n}(\cdot)$. Quasi-randomized Monte Carlo \citep{genz_1992,genz_2009} allows, in fact, accurate calculation of $\Phi_{n}(\cdot)$ for small $n$, and have been recently improved via minimax tilting  \citep{botev_2017} to ensure effective evaluation of $\Phi_{n}(\cdot)$  in moderate $n$ studies. This procedure, available in the \texttt{R} library \texttt{TruncatedNormal}, has a rare vanishing asymptotic relative error, thus allowing tractable inference without sampling from the posterior in studies having, typically, few hundreds of units. This strategy is also useful in larger $n$ applications when few evaluations of $\Phi_{n}(\cdot)$ are required, as in prediction of not many outcomes and in selection among few models. However, for general inferential tasks requiring a plenty of evaluations of  $\Phi_{n}(\cdot)$, such as in numerical integration, moments calculation and high-dimensional prediction or model selection, inference without sampling from the posterior might face non-negligible increments in computational time when $n$ is large; refer to \citet[][\S 5]{botev_2017} for details on scalability in the evaluation of $\Phi_{n}(\cdot)$. In this situation, sampling from the posterior provides a tractable and common strategy to obtain numerical evaluations of generic functionals via Monte Carlo integration approximating $E\{g(\beta) \mid y, X\}=\int g(\beta)\pi(\beta \mid y, X) \mbox{d} \beta$. Indeed, the availability of a large number $R$ of  replicates from the unified skew-normal posterior, allows fast and accurate approximation of $E\{g(\beta) \mid y, X\}$ via $\sum_{r=1}^{\R} g(\beta^{(r)})/R$. 

Popular routines addressing the above goal require data augmentation Markov Chain Monte Carlo \citep[e.g.][]{Albert_1993, holmes_2006, fru_2007, Polson_2013}, which provide poor performance, especially in imbalanced high-dimensional studies \citep{john_2017}. This issue can be addressed via   Algorithm \ref{algo1}, which combines the stochastic representation of the unified skew-normal posterior in Corollary \ref{cor1} with a new scheme proposed by \citet{botev_2017} to obtain independent samples from  multivariate truncated Gaussians. This routine relies on minimax tilting  and accept-reject methods to improve the acceptance rate of classical rejection sampling, while  avoiding convergence and mixing issues of Markov Chain Monte Carlo methods. By combining this sampler with classical routines  for multivariate Gaussians, Algorithm \ref{algo1} inherits these properties, thus improving the computational methods discussed in \S \ref{sec.1}, especially in large $p$  and small-to-moderate $n$ applications. Clearly, when $n$ increases and $p$ decreases, sampling from the $n$-variate truncated Gaussian progressively affects computational time in favor of  more efficient Markov Chain Monte Carlo strategies which directly explore the $p$-dimensional parametric space \citep[e.g.][]{chopin_2017}. In this situation, a possibility to scale-up the computations is to exploit the  structure of Algorithm \ref{algo1} to perform parallel computing.  Another alternative is to leverage the closure properties of multivariate truncated Gaussians under conditioning \citep{Horrace_2005} and iteratively block-update sub-vectors of $V_1$ whose dimension still allows efficient sampling via \citet{botev_2017}. Although this hybrid strategy could induce some auto-correlation in the posterior samples of $\beta$, the blocking approach typically guarantees  improvements in mixing and convergence \citep[e.g.][]{roberts_1997}.

It is also worth noticing that \citet{botev_2017} applied his accept-reject method to Bayesian probit regression. However, unlike  Algorithm \ref{algo1}, the proposed strategy  requires sampling from $(n +p)$-variate truncated Gaussians. Separating these two blocks, as in Algorithm \ref{algo1}, reduces computational complexity and allows parallel computing. A more similar representation can be found in \citet[][\S 2.1]{holmes_2006} and  in the documentation of the \texttt{R} library \texttt{TruncatedNormal} by \citet{botev_2017}. In fact, the resulting routines are closely related to Algorithm \ref{algo1}. However, \citet[][\S 2.1]{holmes_2006} and \citet[][\S 5.4]{botev_2017} base their derivations on different arguments without noticing that the posterior is indeed a unified skew-normal. This last result and its broader implications are arguably the most important contribution of the present article.

\begin{algorithm}[t]
 \caption{Exact scheme to draw independent samples from the posterior in Theorem \ref{teo1} } 
 \label{algo1}
  \For( ){$r$ from $1$ to $R$}
  {\vspace{3pt}
 {[1]}   Sample $V^{(r)}_0$ from $N_{p}\{0_p,\bar{\Omega}-\bar{\Omega}\omega D^{\T}(D \Omega D^{\T}+I_n)^{-1}D \omega\bar{\Omega}\}$. [in \texttt{R} use \texttt{rmvnorm}]
\\
 {[2]}  Sample $V^{(r)}_1$ from an $n$-variate truncated Gaussian with mean vector $0_n$, correlation matrix $s^{-1}(D \Omega D^{\T}+I_n)s^{-1}$ and truncation below $- s^{-1}D\xi$, using the accept-reject algorithm of \citet{botev_2017}.  [in \texttt{R} use \texttt{mvrandn}]
\\
  {[3]}  Compute $\beta^{(r)}$ via $\beta^{(r)}= \xi+\omega\{V^{(r)}_0+ \bar{\Omega}\omega D^{\T}(D \Omega D^{\T}+I_n)^{-1}sV^{(r)}_1\}$\vspace{3pt}}
  {\bf output:} $\beta^{(1)}, \ldots, \beta^{(R)}$
  \vspace{-4pt}
\end{algorithm}

Finally, Algorithm \ref{algo1} can be also adapted to sample from a generic unified skew-normal. This can be broadly useful much beyond Bayesian  inference. An example is parametric bootstrap \citep[e.g.][]{efr_1994} for frequentist inference on  the unified skew-normal parameters.

\subsection{A  class of conjugate unified skew-normal priors for Bayesian probit regression}
\label{sec.25}
The derivations in \S \ref{sec.22} suggest the more general result outlined in Corollary \ref{cor4}, thereby allowing tractable inference in  Bayesian probit regression under more flexible priors for $\beta$.
\begin{corollary}
If $(y_i \mid x_i, \beta) \sim \mbox{\normalfont Bern}\{\Phi(x_i^{\T} \beta)\}$ independently for $i=1, \ldots, n$, and $\beta$ is assigned a $\SUN_{p,m}(\xi,\Omega,\Delta,\gamma,\Gamma)$ prior \citep{arellano_2006}, then
\begin{eqnarray}
(\beta \mid y, X) \sim \SUN_{p,m+n}(\xi_{\post},\Omega_{\post},\Delta_{\post} ,\gamma_{\post}, \Gamma_{\post}),
\label{eq9}
\end{eqnarray}
with posterior parameters $\xi_{\post}=\xi$, $\Omega_{\post}=\Omega$, $\Delta_{\post}=(\Delta  \ \ \bar{\Omega}\omega D^{\T}s^{-1})$, $\gamma_{\post}=(\gamma^{\T}  \ \ \xi^{\T} D^{\T}s^{-1})^{\T}$ and $\Gamma_{\post}$ characterizing an $(m+n) \times (m+n)$ full-rank correlation matrix having block entries $\Gamma_{\post[11]}=\Gamma$, $\Gamma_{\post[22]}=s^{-1}(D \Omega D^{\T}+I_n) s^{-1}$, $\Gamma_{\post[21]}=\Gamma_{\post[12]}^{\T}=s^{-1}D\omega\Delta$.
\label{cor4}
\end{corollary}

According to Corollary \ref{cor4}, tractable inference  in Bayesian probit regression is possible under a broader class of priors. Indeed, all the methods in \S \ref{sec.22}--\S \ref{sec.24} also apply to  this more general case, since  the posterior in  \eqref{eq9} is still a unified skew-normal. This ensures increased flexibility in prior specification, thus allowing departures from normality. Although the general unified skew-normal choice may be  uncommon in applied contexts, it shall be noticed that this class incorporates several  priors of interest, including multivariate Gaussians, independent skew-normals for each $\beta_1, \ldots, \beta_p$, and multivariate skew-normals for $\beta$ \citep{arellano_2006}.

\section{Empirical studies}
\label{sec.3}
To evaluate the methods developed in \S\ref{sec.2} and compare performance with the popular strategies for Bayesian inference in probit regression discussed in \S\ref{sec.1}, let us consider an online available dataset on the gene expression of $n=74$ normal and cancerous biological tissues at $p-1=516$ different tags \citep{mart_2005}. An overarching focus in these applications is quantifying the effects of gene expression on the probability of a cancerous tissue and predicting the status of new tissues as a function of the gene expression \citep[e.g.][]{tzi_2007}. Consistent with this goal, let us focus on studying the location of the posterior for $\beta$  and the predictive distribution in the Bayesian model $(y_i \mid x_i, \beta) \sim \mbox{\normalfont Bern}\{\Phi(x_i^{\T} \beta)\}$, for $i=1, \ldots, n$, with $\beta \sim N_{517}( 0_{517}, 16\cdot I_{517})$ prior. In this probit regression, $x_i$ denotes the vector having an intercept term and the gene expressions for tissue $i$, whereas $y_i$ is either $1$ or $0$ if the tissue is cancerous or not, respectively. 

The choice of a weakly informative prior for the $\beta$ coefficients is motivated by the guidelines in \citet{gelman2008}  and by similar implementations from \citet{botev_2017} and \citet{chopin_2017}.   In line with these contributions, the gene expressions at the  $516$ different tags have been also standardized to have mean $0$ and standard deviation $0{\cdot}5$. To assess predictive performance, the prior for $\beta$ is updated with the information of 50 randomly chosen observations, and out-of-sample classification via the posterior predictive distribution is made on the 24 held-out units.

\begin{table}[b]
\def~{\hphantom{0}}
\tbl{Assessment on computational efficiency. For each sampling scheme under analysis, total number of samples from $(\beta \mid y, X)$ per second and statistics summarizing the effective sample sizes computed from the produced chains for the coefficients $\beta_1, \ldots, \beta_{517}$}{%
\begin{tabular}{lrcrrr}
 \\
& \textsc{samples per second}  && \multicolumn{3}{c}{\textsc{mixing via effective sample sizes}} \\
&  Samples of $\beta$ per second& & Minimum & First quartile & Median \\[2pt]
Unified skew-normal sampler & $886{\cdot}64$ && $20000{\cdot}00$ & $20000{\cdot}00$ & $20000{\cdot}00$ \\
Gibbs sampler & $13{\cdot}48$ && $55{\cdot}46$ & $2417{\cdot}38$ & $3687{\cdot}18$ \\
Hamiltonian no u-turn sampler & $15{\cdot}95$ && $20000{\cdot}00$  & $20000{\cdot}00$  & $20000{\cdot}00$ \\
Adaptive Metropolis--Hastings sampler & $19{\cdot}34$ && $28{\cdot}55$ & $49{\cdot}22$ & $59{\cdot}07$ \\
\end{tabular}}
\label{tab1}
\end{table}

\begin{figure}[t!]
\captionsetup{font={small}}
\includegraphics[height=5.2cm]{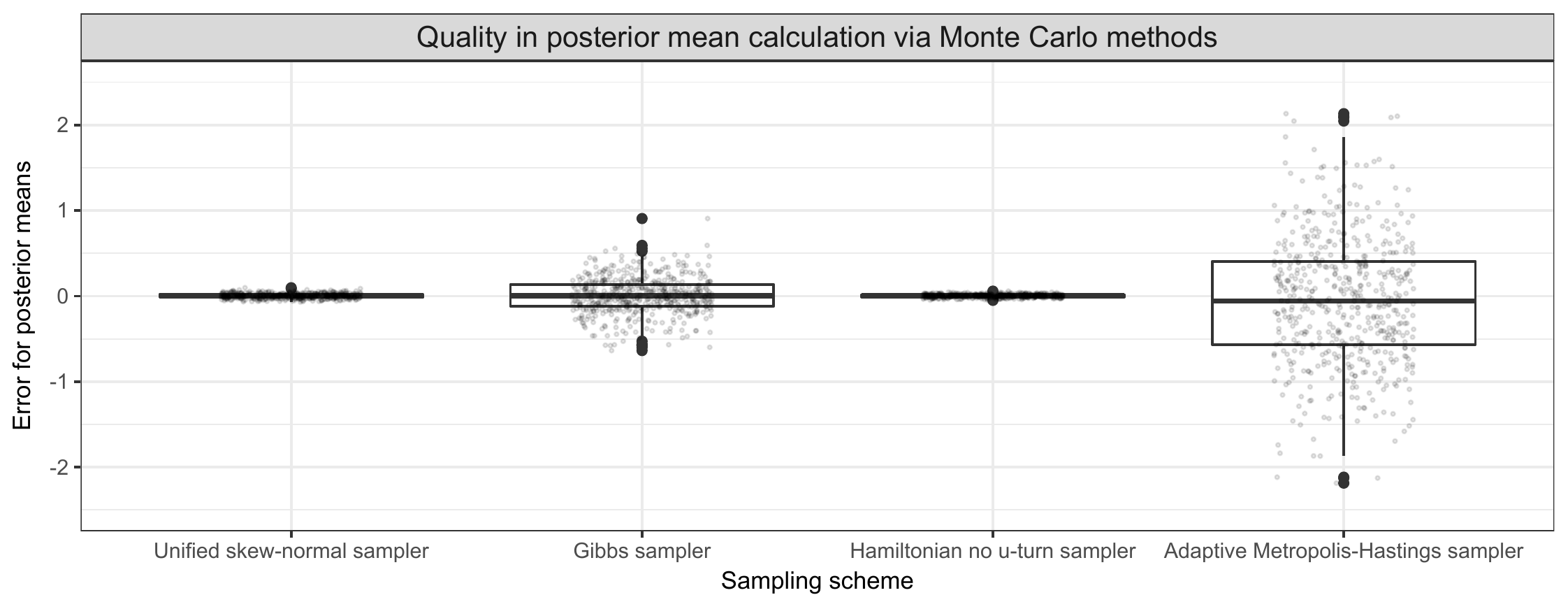}
\caption{Moments calculation performance. For each sampling scheme under study, boxplot of the differences between the posterior means for the coefficients based on the samples from $(\beta \mid y,X)$ and those calculated via \eqref{eq6}. The jittered dots represent the values of the differences from which each boxplot is derived.}
\label{f2}
\end{figure}

\begin{figure}[t]
\captionsetup{font={small}}
\includegraphics[height=5.2cm]{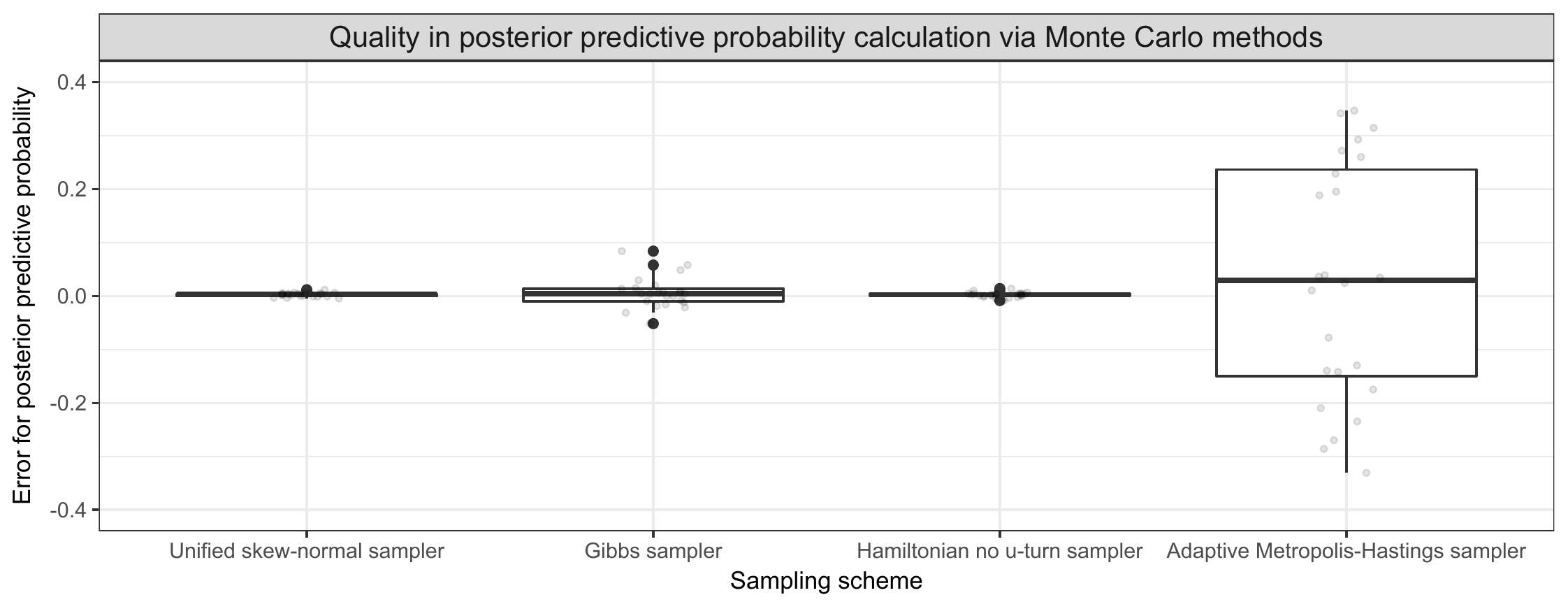}
\caption{Predictive performance. For each sampling scheme under study, boxplot of the differences between the posterior predictive probabilities for the $24$ held-out units based on the samples from $(\beta \mid y,X)$ and those calculated via \eqref{eq7}. The jittered dots represent the values of the differences from which each boxplot is derived.}
\label{f3}
\end{figure}

Although other datasets could be considered, it shall be emphasized that state-of-the-art computational methods for probit regression provide valuable strategies in a variety of applications, but face  mixing and time-inefficiency issues in large $p$ and small $n$ studies  \citep[e.g][]{chopin_2017}. As shown in Figs. \ref{f2}--\ref{f3} and in Table \ref{tab1}, the novel results outlined in \S \ref{sec.2}  allow notable improvements in these large $p$ and small $n$ studies, thus providing straightforward Bayesian inference in relevant applications where this task was previously impractical. To clarify these results, the strategies in  \S \ref{sec.2} are compared with state-of-the-art procedures, covering the data augmentation  Gibbs sampler  by \citet{Albert_1993}, the Hamiltonian no u-turn sampler  in \citet{hoff_2014} and the adaptive  Metropolis--Hastings from \citet{haario_2001}. To increase acceptance rate and efficiency, the starting Gaussian proposal for the Metropolis--Hastings routine has been initialized with the mean and the rescaled variance-covariance matrix provided by an expectation propagation approximation. Consistent with  \citet{chopin_2017} and \citet{rob_2001}, the scaling factor has been set to  $2{\cdot}38^2/p$.

The above Markov Chain Monte Carlo routines were run for $20000$ iterations after a burn-in of $5000$, and can been easily implemented in \texttt{R}, leveraging the libraries  \texttt{bayesm},  \texttt{rstan}, and a combination of \texttt{LaplacesDemon} and \texttt{EPGLM}, respectively. Although certain routines  converged rapidly than others and with excellent mixing, the same  settings were considered for all the algorithms  to  facilitate comparison. The  sampling scheme proposed in Algorithm \ref{algo1} provides instead independent samples from the exact posterior and hence requires no burn-in or convergence checks. Refer to the Supplementary Materials for details on code and implementation. 

According to Table \ref{tab1},  the Hamiltonian no u-turn sampler has the same mixing of Algorithm \ref{algo1} which displays, however, a significantly faster sampling speed. This could be due to the number of leap-frog steps required at each iteration of the no u-turn sampler  \citep{chopin_2017}. As expected, the data augmentation Gibbs sampler and Metropolis--Hastings display lower mixing, but provide similar or improved running time compared to Hamiltonian no u-turn samplers. However, as is clear from Figs. \ref{f2}--\ref{f3},  this reduction in mixing has a direct effect on the accuracy of posterior inference and prediction. There is instead  an almost perfect match between Monte Carlo and direct estimates of posterior means and posterior predictive probabilities for the proposed Algorithm \ref{algo1} and the Hamiltonian no u-turn sampler. However, as already discussed, such a routine is significantly slower than Algorithm \ref{algo1} in this application. These computational gaps further increased when focusing on larger $p$ studies, with the competing methods becoming rapidly impractical. Conversely, the inference and sampling methods relying on the unified skew-normal results have difficulties in scaling with $n$. This claim is confirmed by a voice rehabilitation study presented at the online repository in the Supplementary Materials. However,  also in this application having doubled $n$ and almost halved $p$, Algorithm \ref{algo1} remains still competitive.

\section{Final considerations and future directions of research}
\label{sec.4}
This article shows that the posterior for the coefficients in a probit regression having Gaussian priors is a unified skew-normal  \citep{arellano_2006}, thus allowing key advances in  Bayesian modeling of binary response data, especially for large $p$ and small-to-moderate $n$ studies. Indeed, unified skew-normals have moment generating functions involving known quantities, tractable additive representations, and are closed under marginalization, conditioning and linear transformations, thus facilitating derivations of marginal likelihoods for model selection and posterior predictive distributions. As shown in the empirical assessments in \S \ref{sec.3}, in small-to-moderate $n$ settings with large or even huge $p$, posterior inference can proceed via direct methods or via an efficient sampler from the posterior, which notably improves available computational methods.

The above results could lead to computational gains also in more complex formulations relying on predictor-dependent observed or latent binary data, such as in mixture models for density regression \citep{rodrig_2011}. For instance, leveraging results  in \S \ref{sec.23}, binary classification via Gaussian processes \citep{ras_2006} could avoid sampling or approximations by exploiting closure properties of unified skew-normals, especially under conditioning. Moreover, when binary regression serves as a latent dictionary function, sampling the binary data via \eqref{eq7}, instead of conditioning on $\beta$, could speed-up computations. Finally, the  novel conjugacy results in \S \ref{sec.25}, open new avenues for  incorporation of skewness in prior specification.

There are also different directions for future advances. For instance, improved studies on the moment generating function of the unified skew-normal could facilitate direct calculation of relevant functionals without the need to sample from $(\beta \mid y, X)$.  On the same line, improving the methods for efficient evaluation of $\Phi_n(\cdot)$ in large $n$ applications, either via data transformations, blocking methods \citep{chop_2011} or recent algorithms \citep{gent_2017}, could enlarge the range of applications which allow direct inference, prediction and model selection, without sampling from $(\beta \mid y, X)$. Also approximations of the exact posterior, which preserve the skewness but allow analytical inference provide an interesting direction. Finally, more detailed studies on particular forms of  the prior variance-covariance matrix $\Omega$, such as those associated with $g$-priors or generalized $g$-priors \citep{maruyama_2011} and limiting cases arising from flat priors, could provide novel insights on the effect of these specific choices and, potentially, lead to simplifications in the unified skew-normal parameters which may ease posterior inference. It is also possible to consider hyper-priors for $(\xi, \Omega)$ such as the normal-inverse-Wishart. With this choice, Theorem \ref{teo1} holds only for the full conditional  $(\beta \mid y, X, \xi, \Omega)$. Moreover, it is straightforward to notice that $(\xi, \Omega \mid  y, X, \beta)$ has still a normal-inverse-Wishart kernel, since $(\xi, \Omega)$ only enter the Gaussian prior for $\beta$. Hence, although an hierarchical prior would not allow direct sampling from a unified skew-normal posterior, Gibbs sampling methods can be easily applied to this situation.

As discussed in \S \ref{sec.1}, the availability of an exact posterior with tractable stochastic representations and closure properties can also motivate novel finite-sample and asymptotic theory. Finally, although the studies in \S \ref{sec.3} and the discussion in \S \ref{sec.24} provide the general guidelines on the practical usefulness of the unified-skew results in \S \ref{sec.2}, additional quantitative assessments on scalability and its relations with specific prior settings or dataset structures, are certainly interesting.

\section*{Acknowledgement}
The author is grateful to Adelchi Azzalini, David Dunson and Giacomo Zanella for the stimulating discussions on this topic and the comments received on a first version of this article. The Editor, the Associate Editor and the referees are also acknowledged for their useful suggestions. This article was partially supported by the grant number 670337--INDIMACRO.

\section*{Supplementary materials}
Codes and tutorial implementations for the methods associated with this article are available at \texttt{https://github.com/danieledurante/ProbitSUN}.

\appendix

\appendixone

\section*{Appendix A}
\begin{proof}[of Lemma~\ref{lemma1}]
Let $\Phi(x \beta)^{y}\{1-\Phi(x\beta)\}^{1-y}=\Phi\{(2y-1)x \beta\}$ denote the probability mass function of $y$ in Lemma~\ref{lemma1}. Direct application of the Bayes rule provides 
\begin{eqnarray*}
\pi(\beta \mid y, x) \propto \phi(\beta)\Phi\{(2y-1)x \beta\}=  \phi(\beta)\Phi\{(2y-1)x(x^{2}+1)^{-1/2}\beta; (x^2+1)^{-1} \}.
\end{eqnarray*} 
Hence, letting $\xi_{\post}=0$, $\Omega_{\post}=1$, $\Delta_{\post}=(2y-1)x(x^{2}+1)^{-1/2}$, $\gamma_{\post}=0$,  and $\Gamma_{\post}=(x^2+1)^{-1}+\Delta_{\post}^{\T}\Delta_{\post}=1$, provides the kernel of the unified skew-normal in Lemma \ref{lemma1}, with  correlation matrix $\Omega_{\post}^*$ having block entries $\Omega_{\post[11]}^*=\Omega_{\post[22]}^*=1$ and $\Omega_{\post[21]}^*=\Omega_{\post[12]}^*=\Delta_{\post}$. 
\end{proof}

\begin{proof}[of Theorem~\ref{teo1}]
Adapting the proof of Lemma \ref{lemma1}, it is possible to write the joint probability mass function of the responses $y$  as $\prod_{i=1}^n\Phi\{(2y_i-1)x_i^{\T} \beta\}=\Phi_n(D\beta; I_{n})=\Phi_n\{s^{-1} D\beta;(ss^{\T})^{-1}\}$, with $D$ and $s$ defined as in Theorem~\ref{teo1}.  Combining this likelihood for $y$ with the Gaussian prior for $\beta$ provides 
\begin{eqnarray*}
\pi(\beta \mid y, X) \propto \phi_{p}(\beta-\xi; \Omega)\Phi_n\{s^{-1} D\beta; (ss^{\T})^{-1}\}=\phi_{p}(\beta-\xi; \Omega)\Phi_n\{s^{-1} D\xi +s^{-1} D( \beta-\xi); (ss^{\T})^{-1}\}.
\end{eqnarray*}
To establish the relation among the above kernel and  the unified skew-normal density in \eqref{eq3}, note that $( \beta-\xi)=\omega\bar{\Omega}\bar{\Omega}^{-1} \omega^{-1}( \beta-\xi)$. Therefore, letting $\xi_{\post}=\xi$, $\Omega_{\post}=\Omega$, $\Delta_{\post}=\bar{\Omega}\omega D^{\T} s^{-1}$, $\gamma_{\post}=s^{-1} D\xi$  and $\Gamma_{\post}=(ss^{\T})^{-1}+\Delta_{\post}^{\T}\bar{\Omega}^{-1}\Delta_{\post}=s^{-1}s^{-1}+s^{-1} D\omega\bar{\Omega}\bar{\Omega}^{-1}\bar{\Omega}\omega D^{\T} s^{-1}=s^{-1}(D \Omega D^{\T}+I_n) s^{-1}$, provides the kernel of a unified skew-normal whose parameters coincide with those presented in Theorem ~\ref{teo1}. To conclude the proof it is also necessary to guarantee that 
\[
 \Omega_{\post}^*=\begin{bmatrix}
    s^{-1}(D \Omega D^{\T}+I_n) s^{-1}       & s^{-1} D\omega\bar{\Omega} \\
  \bar{\Omega}\omega D^{\T} s^{-1}     & \bar{\Omega}
\end{bmatrix}=\begin{bmatrix}
    s^{-1} &  0 \\
  0   & \omega^{-1}
\end{bmatrix}  \times \begin{bmatrix}
D \Omega D^{\T}+I_n      & D{\Omega} \\
{\Omega}D^{\T}     & \Omega
\end{bmatrix} \times \begin{bmatrix}
    s^{-1} & 0 \\
  0     & \omega^{-1}
\end{bmatrix}
\]
is a full-rank correlation matrix. This last result can be easily proved by noticing that $\Omega_{\post}^*$ coincides with the correlation matrix of the random vector $(z_1^{\T}, z_2^{\T})^{\T}$ where $z_1=D z_2+ \epsilon$ with $E(\epsilon)=0_n$, $E(\epsilon\epsilon^{\T})=I_n$, and $z_2$ is a $p$-variate random variable having zero mean and positive definite variance-covariance matrix $E(z_2 z_2^{\T})=\Omega=\omega \bar{\Omega}\omega$. Finally,  $s=\mbox{diag}\{(d^{\T}_1\Omega d_1+1)^{1/2}, \ldots, (d^{\T}_n\Omega d_n+1)^{1/2}\}$ is the diagonal matrix with the square root of the diagonal elements of $E(z_1z_1^{\T})=D \Omega D^{\T}+I_n$.
\end{proof}

\begin{proof}[of Corollary~\ref{cor1}]
The proof is a simple adaptation of equation (7.4) in \citet[][\S 7.1.2]{azzalini_2013} to the unified skew-normal posterior in Theorem \ref{teo1}. In particular,  according to  \citet[][\S 7.1.2]{azzalini_2013} the posterior in equation \eqref{eq2} has the same distribution of the random variable
\begin{eqnarray*}
\xi_{\post}+\omega_{\post}(V_0+\Delta_{\post}\Gamma^{-1}_{\post}V_1)
\end{eqnarray*}
with $V_0 \sim N_p(0_p, \bar{\Omega}_{\post}-\Delta_{\post}\Gamma^{-1}_{\post}\Delta_{\post}^{\T})$ and $V_1$ from an $n$-variate truncated normal with mean $0_n$, covariance matrix $\Gamma_{\post}$ and truncation below $- \gamma_{\post}$. Substituting the posterior parameters in this stochastic representation with their expressions in Theorem  \ref{teo1} concludes the proof. To clarify this final claim, note that $\Delta_{\post}\Gamma^{-1}_{\post}= \bar{\Omega}\omega D^{\T}s^{-1}s(D \Omega D^{\T}+{ I}_n)^{-1}s=\bar{\Omega}\omega D^{\T}(D \Omega D^{\T}+{ I}_n)^{-1}s$ and that $\Delta_{\post}\Gamma^{-1}_{\post}\Delta_{\post}^{\T}$ coincides with the matrix $\bar{\Omega}\omega D^{\T}s^{-1}s(D \Omega D^{\T}+{ I}_n)^{-1}ss^{-1}D \omega \bar{\Omega}=\bar{\Omega}\omega D^{\T}(D \Omega D^{\T}+{ I}_n)^{-1}D \omega \bar{\Omega}$.
\end{proof} 

\begin{proof}[of Corollary~\ref{cor2}]
Recalling the expression for $\pi(\beta \mid y, X)$ outlined in equation \eqref{eq3}, the posterior predictive probability $\mbox{pr}(y_{\mbox{\scriptsize new}}=1 \mid y, X, x_{\mbox{\scriptsize new}})=\int\Phi(x_{\mbox{\scriptsize new}}^{\T}\beta)\pi(\beta \mid y, X) \mbox{d} \beta$ can be  expressed as 
\[
\Phi_n\{s^{-1}D\xi ; s^{-1}(D \Omega D^{\T}+I_n) s^{-1} \}^{-1}\int\phi_p(\beta -\xi; \Omega)\Phi(x_{\mbox{\scriptsize new}}^{\T}\beta)\Phi_n\{s^{-1} D\beta; (ss^{\T})^{-1}\} \mbox{d} \beta.
\]
Exploiting the proof of Theorem \ref{teo1}, the quantity inside the above integral can be re-expressed as $\phi_p(\beta -\xi; \Omega)\Phi_{n+1}\{s_{\mbox{\scriptsize  \normalfont  new}}^{-1} D_{\mbox{\scriptsize  \normalfont  new}}\beta; (s_{\mbox{\scriptsize  \normalfont  new}}s_{\mbox{\scriptsize  \normalfont  new}}^{\T})^{-1}\}$, with $D_{\mbox{\scriptsize  \normalfont  new}}$ and $s_{\mbox{\scriptsize  \normalfont  new}}$ defined as in Corollary~\ref{cor2}. Comparing, now, this function with the density of the unified skew-normal posterior in  \eqref{eq3} it can be immediately noticed that $\phi_p(\beta -\xi; \Omega)\Phi_{n+1}\{s_{\mbox{\scriptsize  \normalfont  new}}^{-1} D_{\mbox{\scriptsize  \normalfont  new}}\beta; (s_{\mbox{\scriptsize  \normalfont  new}}s_{\mbox{\scriptsize  \normalfont  new}}^{\T})^{-1}\}$  is the kernel of a unified skew-normal with normalizing constant $\int\phi_p(\beta -\xi; \Omega)\Phi(x_{\mbox{\scriptsize new}}^{\T}\beta)\Phi_n\{s^{-1} D\beta; (ss^{\T})^{-1}\} \mbox{d} \beta=\Phi_{n+1}\{ s_{\mbox{\normalfont  \scriptsize new}}^{-1} D_{\mbox{\scriptsize \normalfont new}}\xi ; s_{\mbox{\scriptsize \normalfont new}}^{-1}( D_{\mbox{\scriptsize \normalfont new}} \Omega D_{\mbox{\scriptsize \normalfont new}}^{\T}+I_{n+1}) s_{\mbox{\scriptsize \normalfont new}}^{-1}\}$. To conclude the proof, substitute this quantity in the above formula for the posterior predictive probability $\mbox{pr}(y_{\mbox{\scriptsize new}}=1 \mid y, X, x_{\mbox{\scriptsize new}})$. 
\end{proof}

\begin{proof}[of Corollary~\ref{cor3}]
To prove Corollary~\ref{cor3} simply notice that  $\int \mbox{\normalfont pr}(y {\mid}  \mathcal{M}_k, X, \beta_{\mathcal{J}_k})\pi( \beta_{\mathcal{J}_k} {\mid} \mathcal{M}_k) \mbox{d}\beta_{\mathcal{J}_k}$ is the normalizing constant of the posterior for $\beta_{\mathcal{J}_k}$ in model $\mathcal{M}_k$. Hence, adapting equation \eqref{eq3} to model $\mathcal{M}_k$, leads to $\int \mbox{\normalfont pr}(y \mid  \mathcal{M}_k, X, \beta_{\mathcal{J}_k})\pi( \beta_{\mathcal{J}_k} \mid \mathcal{M}_k) \mbox{d}\beta_{\mathcal{J}_k}=\Phi_n\{s_k^{-1}D_k\xi_k;s_k^{-1}(D_k \Omega_k D_k^{\T}+I_n) s_k^{-1} \}$.
\end{proof}

\begin{proof}[of Corollary~\ref{cor4}]
To prove Corollary \ref{cor4} it suffices to generalize Theorem \ref{teo1}. In particular, adapting the proof of Theorem \ref{teo1} to the case in which $\beta\sim \SUN_{p,m}(\xi,\Omega,\Delta,\gamma,\Gamma)$, provides 
\[
\pi(\beta \mid y, X) \propto \phi_{p}(\beta-\xi, \Omega)\Phi_n\{s^{-1} D\beta; (s s^{\T})^{-1}\}\Phi_m\{\gamma+\Delta^\T \bar{\Omega}^{-1} \omega^{-1}(\beta-\xi); \Gamma-\Delta^{\T}\bar{\Omega}^{-1}\Delta \}.
\]
To proceed with the proof, note that exploiting Theorem \ref{teo1}, it is possible to re-write $\Phi_n\{s^{-1} D\beta; (s s^{\T})^{-1}\}$ as $\Phi_n\{s^{-1} D\xi +(\bar{\Omega}\omega D^{\T} s^{-1})^{\T}\bar{\Omega}^{-1} \omega^{-1}( \beta-\xi); s^{-1}(D \Omega D^{\T}+I_n) s^{-1}- s^{-1} D\omega\bar{\Omega}\bar{\Omega}^{-1}\bar{\Omega}\omega D^{\T} s^{-1} \}.$ 

Consistent with this result, let us define the appropriate parameters for which the above kernel coincides with the one of the unified skew-normal posterior in Corollary~\ref{cor4}. This goal is easily accomplished by setting $\xi_{\post}=\xi$, $\Omega_{\post}=\Omega$,  $\Delta_{\post}=(\Delta \ \ \bar{\Omega}\omega D^{\T} s^{-1})$, $\gamma_{\post}=(\gamma^{\T}  \ \  \xi^{\T} D^{\T}s^{-1})^{\T}$ and $\Gamma_{\post}$ a full-rank correlation matrix with blocks $\Gamma_{\post[11]}=\Gamma$, $\Gamma_{\post[22]}=s^{-1}(D \Omega D^{\T}+I_n) s^{-1}$, $\Gamma_{\post[21]}=\Gamma_{\post[12]}^{\T}=s^{-1}D\omega\Delta$. As in Theorem \ref{teo1}, it is also necessary to ensure that 
\[
 \begin{bmatrix}
 \Gamma &  \Delta^{\T}\omega D^{\T} s^{-1}      & \Delta^{\T} & \\
s^{-1} D\omega \Delta & \  \ s^{-1}(D \Omega D^{\T}+I_n) s^{-1}   \ \    & s^{-1} D\omega\bar{\Omega}  \\
\Delta &  \bar{\Omega}\omega D^{\T} s^{-1}     & \bar{\Omega}
\end{bmatrix},
\]
is a full-rank correlation matrix. This result follows under minor modifications of the proof in Theorem \ref{teo1}, after noticing that by definition the $(m+p) \times (m+p)$ unified skew-normal prior matrix $\Omega^*$  having block entries $\Omega_{[11]}^*=\Gamma$, $\Omega_{[22]}^*=\bar{\Omega}=\omega^{-1}\Omega\omega^{-1}$, $\Omega_{[21]}^*=\Omega_{[12]}^{*\T}=\Delta$, is a full-rank correlation matrix.
\end{proof}
\bibliographystyle{biometrika}
\bibliography{paper-ref}

\end{document}